\documentclass[10pt,aps,prd,nofootinbib,reprint,groupedaddress,twocolumn]{revtex4-1}

\usepackage[utf8]{inputenc}
\usepackage{amsmath}
\usepackage{graphicx}
\usepackage{amssymb}
\usepackage{bbold}
\usepackage{physics}
\usepackage{hyperref}
\usepackage{subcaption}
\usepackage{comment}
\hypersetup{colorlinks=true,allcolors=[rgb]{0.5,0.0,0.5}}

\newcommand{\orcid}[1]{\href{https://orcid.org/#1}{#1}}

\newcommand{\beq}{\begin{equation}}
\newcommand{\eeq}{\end{equation}}
\newcommand{\bea}{\begin{eqnarray}}
\newcommand{\eea}{\end{eqnarray}}
\newcommand{\eps}{\varepsilon}

\begin{document}

\title{Testing the dark side of neutrino oscillations with the  solar neutrino fog \\at Dark Matter experiments}

\author{Julia Gehrlein}
\email{julia.gehrlein@colostate.edu}
\thanks{\orcid{0000-0002-1235-0505}}
\affiliation{Physics Department, Colorado State University, Fort Collins, CO 80523, USA}
\author{Tanmay Kushwaha}
\email{tanmay.kushwaha@colostate.edu}
\thanks{\orcid{0000-0002-9847-1278}}
\affiliation{Physics Department, Colorado State University, Fort Collins, CO 80523, USA}

\begin{abstract}
The recent detection of the solar neutrino background at Dark Matter direct detection experiments paves the way to fully explore an important degeneracy in neutrino oscillations in the presence of new interactions, named the LMA-Dark degeneracy. This degeneracy makes it impossible to determine the neutrino mass ordering in  oscillation experiments if neutrinos have  new vectorial interactions with matter. As the composition of solar neutrinos at the Earth consists of all three neutrino flavors, testing the presence of new neutrino interactions  in the muon and tau neutrino sector in  scatterings can fully probe the LMA-Dark region for the first time.
In this paper we show that current data from XENONnT and PandaX-4T does not yet 
exclude the LMA-Dark region with equal couplings of a new mediator to muon and tau neutrinos and quarks, and we identify the possible experimental scenarios to do so in the future. We also show that Dark Matter experiments can distinguish new interactions in the muon or tau sector only from new interactions affecting both sectors.  

\end{abstract}

\maketitle

\section{Introduction}
Since its prediction in 1974 \cite{Freedman:1973yd,Drukier:1984vhf} coherent elastic neutrino nucleus scattering (CEvNS) has been observed with several detector materials like CsI,  argon, and germanium using neutrinos produced in pion decay at rest \cite{COHERENT:2017ipa, COHERENT:2020iec,COHERENT:2024axu,COHERENT:2025vuz} as well as using reactor neutrinos with  germanium detectors
\cite{Colaresi:2022obx,Colaresi:2021kus,
Ackermann:2025obx}. 

Recently, the first observations
of CEvNS with solar neutrinos at dark matter (DM) direct detection experiments has been reported by 
 the large-scale xenon experiments XENONnT and PandaX-4T with more than $2.5\sigma$ significance \cite{XENON:2024ijk,PandaX:2024muv}.
This so-called ``neutrino fog" originates from $^8$B solar neutrinos and provides  an irreducible background at dark matter direct detection experiments \cite{Monroe:2007xp,Vergados:2008jp,Strigari:2009bq,Gutlein:2010tq,Billard:2013qya,OHare:2021utq,Akerib:2022ort}. 
 In the Standard Model (SM), the CEvNS process is mediated by the $Z-$boson however, in scenarios beyond the SM where neutrinos have nonstandard interactions (NSI) new contributions to the CEvNS cross sections can arise. In
\cite{Dutta:2017nht,Boehm:2018sux,Schwemberger:2022fjl,Amaral:2023tbs} the effect of  new interactions on the neutrino fog has been studied
and the PandaX-4T and XENONnT data sets have been analyzed by the neutrino community to probe new neutrino interactions in \cite{AristizabalSierra:2024nwf,Li:2024iij,Blanco-Mas:2024ale,Maity:2024aji,Demirci:2024vzk,DeRomeri:2024hvc,Karadag:2025muq,DeRomeri:2024dbv,DeRomeri:2024iaw}.

Here we point out that a detection of the solar neutrino fog can provide crucial constraints on a particular region in the parameter space of new neutrino interactions called the LMA-Dark degeneracy \cite{deGouvea:2000pqg,Bakhti:2014pva,Coloma:2016gei,Denton:2021vtf}. In the LMA-Dark scenario, neutrinos have a new vectorial interaction with SM fermions with a strength of the order of the weak interaction. Probing this large interaction, however, is challenging as the LMA-Dark point is an exact degeneracy of the Hamiltonian which governs neutrino oscillations in matter in the presence of new interactions. Therefore, we have to resort to low energy neutrino scattering processes, which do not suffer from this degeneracy, to test this point in parameter space. As we show in the following, the neutrino fog is the ideal process to achieve this goal.

Probing the LMA-Dark degeneracy is essential as its presence
makes it impossible to determine the neutrino mass ordering at long-baseline experiments and can therefore prevent upcoming experiments like DUNE \cite{DUNE:2020ypp}, HyperKamiokande \cite{Hyper-Kamiokande:2018ofw}, and JUNO \cite{JUNO:2022mxj} to achieve one of their major goals. The open question of the neutrino mass ordering is equivalent to determining the sign of the mass splitting $\Delta m_{31}^2=m_3^2-m_1^2$.\footnote{Here we make the choice of defining  the mass eigenstates based on $|U_{e1}|>|U_{e2}|>|U_{e3}|$ which leads to the sign of $\Delta m_{21}^2$ to be determined experimentally, see \cite{Denton:2020exu,Denton:2021vtf}.} If $\Delta m_{31}^2>0$ then the neutrino mass ordering is normal (NO) with $m_1<m_2<m_3$, if $\Delta m_{31}^2<0$ then the neutrino mass ordering is inverted with $m_3<m_1<m_2$ (IO). From current global fit neutrino oscillation data \cite{Esteban:2024eli} there is a mild  preference for normal ordering however  individual data sets have shown slight preference for different orderings leading to a  mild tension between them \cite{Kelly:2020fkv,Denton:2020uda,Chatterjee:2020kkm,Esteban:2024eli,Chatterjee:2024kbn}. Therefore, the question of the mass ordering currently remains unanswered by oscillation experiments.

Apart from addressing the question of how ``normal" neutrinos are and determining one of the last unknown parameters in the three-flavor neutrino paradigm \cite{Denton:2022een}, determining the neutrino mass ordering has  important implications for other experiments, in particular for experiments which measure observables related to the absolute neutrino mass scale.
For example, the predicted parameter space for the beta decay end point spectrum observable is lower for NO than for IO, affecting the reach of the KATRIN experiment \cite{KATRIN:2001ttj} and future experiments \cite{Monreal:2009za,Gastaldo:2017edk,Alpert:2014lfa}.
Current cosmological observatories like DESI \cite{DESI:2016fyo}, the Atacama Cosmology Telescope \cite{ACT:2025tim}, and the South Pole Telescope \cite{SPT-3G:2025bzu} are close to measuring the sum of the neutrino masses; in fact, the strongest bounds from DESI are in conflict with the minimal allowed value in IO \cite{DESI:2025ejh}. Unambiguously determining the neutrino mass ordering from oscillation experiments will be essential to determine if there is possible new physics in cosmology or even in oscillation physics. Also the detection prospects of the cosmic neutrino background by capture on tritium depends on the mass ordering
\cite{Long:2014zva}.
Finally, the allowed parameter space for neutrinoless double beta decay experiments allows much smaller minimal values of $|m_{\beta\beta}|$ depending on the lightest mass in NO than in IO by at least one order of magnitude in the hierarchical region of neutrino masses  \cite{Denton:2023hkx}. Correctly identifying  the mass ordering is therefore essential for neutrinoless double beta decay experiments in determining their required sensitivity.

In the following we first review the LMA-Dark degeneracy and its current status in sec.~\ref{sec:LMAD} (see \cite{Denton:2022nol} for a more detailed overview). In sec.~\ref{sec:nuscat} we discuss neutrino scattering at DM direct detection experiments before  we 
outline our analysis in sec.~\ref{sec:analysis}. Our  results and next steps in fully ruling  out the LMA-Dark degeneracy are described in sec.~\ref{sec:results} and  we conclude in sec.~\ref{sec:conclusion}.

\section{LMA-Dark degeneracy}
\label{sec:LMAD}
New neutral current vectorial neutrino interactions with matter fermions can be parametrized in an effective field theory framework as four-fermion operators \cite{Wolfenstein:1977ue,Proceedings:2019qno}
\begin{align}
\mathcal L_{\rm NSI}=-2\sqrt2G_F\sum_{f,\alpha}\eps^{f,V}_{\alpha\beta}(\bar\nu_\alpha\gamma^\mu P_L\nu_\beta)(\bar f\gamma_\mu f)\,,
\label{eq:lag}
\end{align}
where $\varepsilon_{\alpha\beta}^{f,V}$ are the NSI parameters which parametrize the strength of the new interaction relative to the Fermi constant $G_F$ between neutrino flavors $\alpha,~\beta$ and matter fermions $f=e,~u,~d$.
The off-diagonal NSI parameters can be complex while the flavor diagonal couplings are real.
These new interactions arise in a variety of models \cite{Heeck:2011wj, Farzan:2015doa,Farzan:2015hkd, Farzan:2016wym,Forero:2016ghr,Babu:2017olk, Heeck:2018nzc,Denton:2018dqq,Dey:2018yht, Babu:2019mfe, Babu:2020nna, Babu:2021cxe}.

NSI can be probed via either neutrino oscillations where they appear as a new matter effect or by neutrino scattering processes where they lead to a new contribution to the cross section. 
In oscillations the new interactions appear in the Hamiltonian which describes neutrino oscillations in matter 
\begin{multline}
H=\frac1{2E}\left[U
\begin{pmatrix}
0&0&0\\
0&\Delta m^2_{21}&0\\
0&0&\Delta m^2_{31}
\end{pmatrix}
U^\dagger+\right.\\
\left.a
\begin{pmatrix}
1+\eps_{ee}&\eps_{e\mu}&\eps_{e\tau}\\
\eps_{e\mu}^*&\eps_{\mu\mu}&\eps_{\mu\tau}\\
\eps_{e\tau}^*&\eps_{\mu\tau}^*&\eps_{\tau\tau}
\end{pmatrix}
\right]
\,,
\label{eq:hamiltonian}
\end{multline}
where $E$ is the neutrino energy,
 $U\equiv R_{23}(\theta_{23})U_{13}(\theta_{13},\delta)R_{12}(\theta_{12})$ is the PMNS mixing matrix \cite{Pontecorvo:1957cp,Maki:1962mu} that is parameterized in the usual way \cite{ParticleDataGroup:2024cfk}, $a\equiv2\sqrt2G_FN_eE$ is related to the Wolfenstein matter potential \cite{Wolfenstein:1977ue}, $G_F$ is Fermi's constant, and $N_e$ is the electron number density. In the presence of NSI all entries in the second matrix can be populated.
The NSI parameters in the Hamiltonian are related to the NSI parameters in the Lagrangian from eq.~\eqref{eq:lag} as
\begin{equation}
\eps_{\alpha\beta}=\sum_{f\in\{e,u,d\}}\frac{N_f}{N_e}\eps_{\alpha\beta}^{f,V}\,,
\end{equation}
with the fermion matter density $N_f$.

Constraints on NSI from oscillations have been derived in \cite{Esteban:2024eli} and no strong evidence for nonzero NSI has been found.
However, the oscillation probability derived from eq.~\eqref{eq:hamiltonian} suffers from a degeneracy in the presence of NSI, called the LMA-Dark degeneracy \cite{deGouvea:2000pqg,Bakhti:2014pva,Coloma:2016gei,Denton:2021vtf}.
Namely, as a consequence of CPT invariance, the  Hamiltonian transforms as $H_{\rm vac}\to-H_{\rm vac}^*$, which corresponds to a change in the oscillation parameters as 
\begin{equation}
\begin{gathered}
\Delta m^2_{21}\to-\Delta m^2_{21}\,,\quad
\Delta m^2_{31}\to-\Delta m^2_{31}\,,\\
\delta\to-\delta\, 
\end{gathered}
\end{equation}
and a  simultaneous change in NSI parameters parametrized 
via the parameter $x$
as \cite{Denton:2018xmq}
\begin{equation}
(\eps_{ee},~\eps_{\mu\mu},~\eps_{\tau\tau})=(x-2,~x,~x)\, 
\label{eq:x}
\end{equation}
with $\eps_{\alpha\beta}=0$ for $\alpha \neq\beta$.\footnote{Depending on the definition of the mass eigenstates, the LMA-Dark degeneracy can also be defined as $\theta_{12}>45^\circ$ instead of $\Delta m_{21}^2<0$ \cite{Denton:2020exu, Denton:2021vtf}.}
If $x=0$ then the LMA-Dark degeneracy is fully contained in the electron neutrino sector, if $x=2$ the LMA-Dark degeneracy is in the muon and tau neutrino sector.

In turn, the LMA-Dark degeneracy also means that one cannot measure the mass ordering unambiguously with oscillation experiments.
Here an alternative way to test NSI comes to the rescue, namely neutrino neutral current scattering experiments.
These experiments do not suffer from the LMA-Dark degeneracy but have their own degeneracies, unrelated to LMA-Dark. Since 
matter effects in oscillations are coherent forward scattering processes, they are independent of the mediator mass of the new interaction. 
In contrast, in scattering processes the mediator mass plays a role. Therefore, to probe the LMA-Dark degeneracy one has to test all mediators down to light masses. Indeed,  cosmology excluded neutrinophilic mediators with mass below the MeV scale. For a vector mediator the bound reads $m_{Z'}\gtrsim 3 $ MeV, depending on the data sets combined \cite{Kamada:2015era,Huang:2017egl,Blinov:2019gcj,Sabti:2021reh,Li:2023puz,Ghosh:2024cxi}.\footnote{Our model also predicts neutrino self-interactions due to the new mediator which can be constrained using supernova data  \cite{Fiorillo:2022cdq,Fiorillo:2023cas,Fiorillo:2023ytr,Akita:2022etk,Telalovic:2024cot}. }
So far, the LMA-Dark region has been excluded for a mediator with masses above the GeV scale with couplings to quarks and electrons using 
NuTeV and CHARM data where neutral current deep inelastic scattering is dominant \cite{Coloma:2017ncl,Coloma:2017egw}. At lower energies the dominant neutral current process is CEvNS.
Using CEvNS data sets from COHERENT, the LMA-Dark region with mediator masses above 50 MeV and coupling to  quarks and electron  or muon neutrinos has been excluded  in \cite{Denton:2018xmq,Chaves:2021pey}. For lighter mediators, 
 CEvNS data from the reactor experiment Dresden-II  has excluded the LMA-Dark region for couplings between electron neutrinos and quarks \cite{Denton:2022nol}. Additionally recent data from CONUS+ \cite{CONUSCollaboration:2024kvo,DeRomeri:2025csu,Chattaraj:2025fvx} excluded this parameter space as well, see sec.~\ref{sec:results}.
 The LMA-Dark parameter space for the coupling of electron neutrinos to electrons down to mediator masses constrained by cosmology has been excluded by  solar neutrino experiments like Borexino 
\cite{Dutta:2020che,BOREXINO:2018ohr}, and reactor experiments like TEXONO, CHARM-II, and GEMMA \cite{Lindner:2018kjo,TEXONO:2009knm,CHARM-II:1994dzw,Beda:2010hk}.  
As there are currently no constraints on the LMA-Dark parameter space for the coupling of muon and tau neutrinos to electrons or quarks, this remains the only
 viable region for LMA-Dark.\footnote{Very specific combinations of  couplings of electron neutrinos to up and down quarks are allowed as well \cite{Denton:2022nol}.}

There are two possible  ways to probe LMA-Dark in the muon and tau neutrino sector: Low threshold experiments using neutrinos from pion decay at rest sources  like Coherent CAPTAIN Mills \cite{CCM:2021leg},  or at the
ESS \cite{Baxter:2019mcx,Chattaraj:2025rtj} or SNS \cite{COHERENT:2024axu, COHERENT:2025vuz} can constrain directly the coupling of a new mediator to muon neutrinos and quarks \cite{Shoemaker:2021hvm,Chaves:2021pey} while the muon neutrino-electron elastic scattering cross section is several orders of magnitude smaller.
Combined with information from oscillation experiments which allow only for a small difference in the new interactions of muons and tau neutrinos
with quarks $|\varepsilon_{\mu\mu}^{u,d}-\varepsilon_{\tau\tau}^{u,d}|\lesssim 2\times 10^{-2}$  
\cite{Coloma:2023ixt} constraints on new interactions in the tau sector can be derived.
Here we  follow an alternative route which allows to directly probe new interactions of muon and tau neutrinos simultaneously, namely the observation of  solar neutrinos in a low energy scattering processes.  Since solar neutrinos arrive at the Earth dominantly as mass eigenstate $\nu_2$ they contain a large fraction of the muon neutrino and tau neutrino flavor eigenstates \cite{Wolfenstein:1977ue}.
A study of the couplings of muon and tau neutrinos to electrons is possible using solar neutrino-electron scattering data \cite{Coloma:2022umy,Brdar:2023ttb,Kelly:2024tvh,Mishra:2023jlq} like current Borexino data \cite{BOREXINO:2022abl} and future JUNO data \cite{JUNO:2023zty} and DUNE  data \cite{Capozzi:2018dat}
as well as neutrino-electron scattering in DM direct detection experiments.
Here we will focus on testing the new interactions of muon and tau neutrinos with quarks using the CEvNS process instead.

Note that SNO's observation of   solar neutrino neutral current scattering on deuterium \cite{SNO:2002tuh,SNO:2011hxd} is dominated by the axial-vector current whereas the LMA-Dark degeneracy appears for the vector current only. Therefore we need to resort to low energy scattering processes  with a vector current like CEvNS.
In the next section we analyze the $^8$B solar neutrino scattering data from PandaX-4T and XENONnT and show that current data does not exclude LMA-Dark yet and outline possible future experimental configurations to test LMA-Dark.

\section{Neutrino scattering at DM direct detection experiments}
\label{sec:nuscat}
The dominant contribution to the solar neutrino fog is neutrinos originating from the 
process $^8\text{B}\to \text{Be}^*+e^++\nu_e$ in the core of the Sun.
Neutrinos from the  $hep$ process have similar energies but  their flux  is three orders of magnitude smaller  than for $^8$B neutrinos \cite{8BFluxNormalization} and hence do not contribute significantly to the neutrino fog. 
After their production,  due to the high matter densities in the Sun and  matter effects, the $^8$B neutrinos undergo 
an adiabatic flavor transition where they propagate as an asymptotic propagation eigenstate $\nu_2$ before they exit the Sun,  as shown in the fundamental paper by Wolfenstein \cite{Wolfenstein:1977ue}.
Since this eigenstate contains all three neutrino flavor eigenstates,
 this means that $^8$B neutrinos provide a source of low energy muon and tau neutrinos without the need to rely on meson, muon, or tau decays.

\begin{figure}
    \centering    
\includegraphics[width=\linewidth]{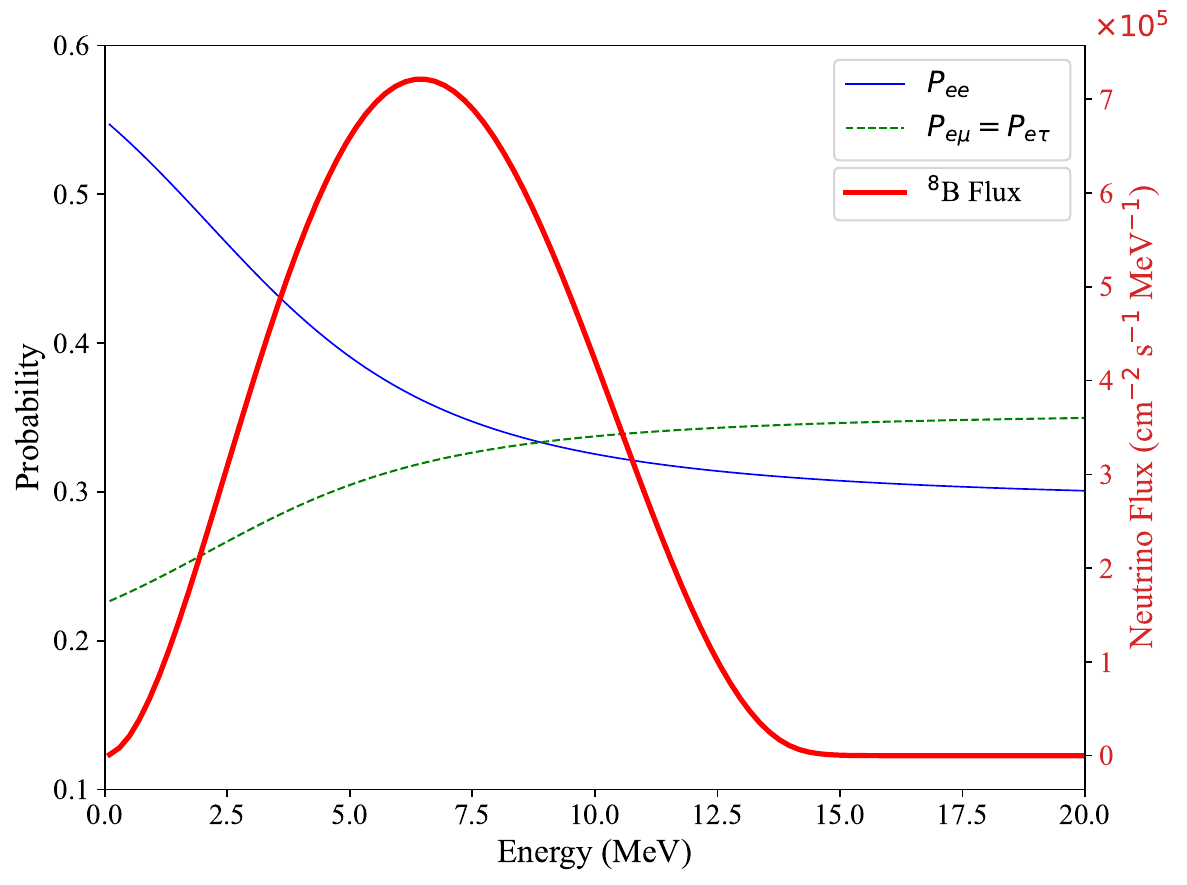} 
        \caption{Solar neutrino survival and transition probability for each neutrino flavor calculated using the PEANUTS code \cite{PEANUTS}, and flux of solar $^8$B neutrinos  \cite{8BSpectrumShape, 8BFluxNormalization,bahcall_b8_spectrum} (thick line). For concreteness we have used as input parameters $\sin\theta_{12}=0.55$, $\sin\theta_{13}=0.15$, $\sin^2\theta_{23}=0.50$, and $\cos \delta=0$ leading to identical probabilities $P_{e\mu}$ and $P_{e\tau}$ shown in dashed.} 
         \label{fig:osci_prob}
\end{figure}

 In the presence of NSI, in general, the flavor transition probability in the Sun gets affected. However,  the LMA-Dark parameter space in the Sun does not change the oscillation probability. Therefore, we do not include the effects of NSI in the propagation of the neutrinos through the Sun. 
 In fig.~\ref{fig:osci_prob} we show the 
 neutrino survival and transition probabilities for $^8$B neutrinos  using the PEANUTS code \cite{PEANUTS}, along with the flux of $^8$B neutrinos. The shape of the $^8$B spectrum was obtained from \cite{8BSpectrumShape, bahcall_b8_spectrum}, and the normalization was obtained from the B16-GS98 Standard Solar Model \cite{8BFluxNormalization}. While the survival probability for electron neutrinos is  accurately known, the
individual transition probabilities $P_{e\mu}$ and $P_{e\tau}$  depend  sensitively on $\sin^2 \theta_{23}$ and  $\cos\delta$ which are currently the least well measured parameters of the leptonic mixing matrix \cite{Esteban:2024eli}. 
For concreteness we use the following values for the parameters of the PMNS matrix
$\sin\theta_{12}=0.55$, $\sin\theta_{13}=0.15$, $\sin^2\theta_{23}=0.50$, and $\cos \delta=0$.
For this set of parameters  the transition probabilities $P_{e\mu}$ and $P_{e\tau}$ are identical.

Hence, the dominant effect of the LMA-Dark NSI parameter space is in the detection of the neutrinos.
The total differential event rate for CEvNS is  given by the sum of the rates from each flavor of neutrinos like
\begin{align}
\frac{dR}{dE_\text{N}} = \frac{dR_e}{dE_\text{N}} + \frac{dR_\mu}{dE_\text{N}} + \frac{dR_\tau}{dE_\text{N}}
\end{align}
where the event rate for $\nu_\alpha$ is given by
\begin{align}
\frac{dR_\alpha}{dE_\text{N}} = \int\ P_{e\alpha}(E_\nu)\frac{d\phi}{dE_\nu} \frac{d\sigma_{\nu_\alpha N}}{dE_\text{N}} dE_\nu
\end{align}
Here $E_\nu$ is the incoming neutrino energy, and $E_N$ is the nuclear recoil energy. $P_{e\alpha}(E_\nu)$ denotes the energy dependent probability of observing a neutrino of flavor $\alpha$ on Earth originating from the $^8$B solar neutrino flux $\phi$. 
The differential cross section for CEvNS is given by \cite{Freedman:1973yd}
\begin{align}
\frac{d\sigma_{\nu_\alpha N}}{dE_\text{N}}=\frac{G_F^2}{\pi} \frac{Q_{w\alpha}^2}{4} \text{ }M_{\text{N}} \left( 1 - \frac{M_{\text{N}} E_\text{N}}{2E_\nu^2} \right) ~F^2(q^2)
\end{align}
where $M_{\text{N}}$ is the target nuclei mass, which for the xenon nucleus, the target material of PandaX-4T and XENONnT,  is 131.293 u,  calculated as a weighted average across their isotopes. $F(q^2)$ is the form factor, for which we use the Helm parametrization as a function of the Bessel function $j_1$  given by \cite{Helm:1956zz}
\begin{align}
F(q^2) = 3 \cdot \frac{j_1(q R_0)}{q R_0} \cdot \exp\left(-\frac{1}{2} (q s)^2\right)
\end{align}
The diffraction radius $R_0$ is calculated using a folding width $s = 0.9$ fm \cite{LEWIN199687} and the rms charge radius of xenon taken from \cite{ANGELI201369}.
Finally, $Q_{w\alpha}^2$ is the weak charge in the presence of NSI. For the case of only diagonal nonzero NSI, it is given by
\begin{align}
    \frac{Q_{w\alpha}^2}{4} = \left[ Z \left( g_p + 2\varepsilon_{\alpha\alpha}^u + \varepsilon_{\alpha\alpha}^d \right) \right. \nonumber\\+\left. N \left( g_n + \varepsilon_{\alpha\alpha}^u + 2\varepsilon_{\alpha\alpha}^d \right) \right]^2
    \label{eq:Qweak}
\end{align}
 Here, $Z=54$ and $N=77.293$ are the proton and neutron numbers of the xenon target nuclei, and  $g_n = - 0.5$ and $g_p = 0.5 - 2 \sin^2{\theta_W}$, where $\sin^2{\theta_W} = 0.2387$ \cite{ParticleDataGroup:2024cfk} are the SM couplings of the $Z-$boson.

For light mediators whose mass $m_{Z'}$ is below the  momentum transfer of the scattering  $q=\sqrt{2 m_A E_r}\lesssim \mathcal{O}(10 $ MeV) eq.~\eqref{eq:Qweak} we have to replace\footnote{We note that in some references
a factor of 2 appears in the denominator of this expression, however as demonstrated in the   appendix of \cite{AtzoriCorona:2022moj} this factor is erroneous.}
\begin{align}
    \varepsilon_{\alpha\alpha}^f = \frac{g_{\nu_\alpha}g_f}{\sqrt{2} \, G_F \, (q^2 + M_{Z'}^2)} = \varepsilon^f_{\alpha\alpha}(q^2=0)\frac{M_{Z'}^2}{(q^2 + M_{Z'}^2)} ~.
\end{align}
where $\varepsilon^f_{\alpha\alpha}(q^2=0)$ is the NSI parameter relevant for oscillations at $q^2=0$ and $g_{\nu_\alpha}$ and $g_f$ are the couplings of the new mediator to neutrinos of flavor $\alpha$ and matter fermions $f$.

\section{Analysis and Results}
\label{sec:analysis}
We turn now to our analysis of the data from XENONnT \cite{XENON:2024wpa} and PandaX-4T \cite{PandaX-4T:2021bab}. Both experiments are dual-phase time projection chambers featuring 5.9 t and 3.7 t of liquid xenon as targets. XENONnT across its two runs SR0 and SR1, with a live time of 108 days and 208.5 days respectively, reported the observation of 37 total events with a background expectation of 26.4$^{+1.4}_{-1.3}$, leading to an observation of  $^8$B solar neutrinos at 2.73$\sigma$. The experiment has a threshold of 0.5 keV and a 3.51 t-yr exposure across both runs. 
PandaX-4T also presented results from two runs, the commissioning run, and the first science run of the experiment. PandaX-4T did two separate analyses for their $^8$B CEvNS data, one where they required coincidence of S1 and S2 signals (paired), and another where they used ionization signals only (US2).
We only use the US2 data in this analysis due to the higher statistics. PandaX-4T reports observing 75 $^8$B neutrino events in their US2 data with a total exposure of 1.04 t-yr. 

XENONnT presents the results in cS2 bins, which are the corrected number of photoelectrons (PE) from the ionization signal. The electron gain $g_2$ = 16.9 PE/electron \cite{XENON:2024ijk} for XENONnT relates the number of ionization electrons released to the number of PE observed in the cS2 signal. The event rate in each electron/cS2 bin is calculated using a Poisson smearing of events.  
The number of events in the $i^{th}$ bin is
\begin{align}
R^i = \mathcal{E}^{(X,P)} \int P(i) ~\epsilon^{X,P}(E_N) ~\frac{dR}{d E_N} \, d E_N
\end{align}
where $P(i)$ is the Poisson probability of getting $i$ electrons/S2
\begin{align}
    P(i) = (n^{X,P})^i \frac{\exp(-n^{X,P})}{\Gamma(i+1)}
\end{align}
when $n^{X,P}$ is the mean number of electrons/PE released for an event with nuclear recoil energy $E_N$ for PandaX-4T/XENONnT respectively, calculated using the charge yields from  \cite{PandaX:2022aac, XENON:2024xgd}. 
$\mathcal{E}^{X,P}$ is the exposure of the respective experiment, with $\mathcal{E}^X = 3.51$ t-yr for XENONnT, and $\mathcal{E}^P = 1.04$ t-yr for PandaX-4T US2. $\epsilon^{X,P}$ is the efficiency of the experiments taken from \cite{PandaX:2024muv} for PandaX-4T and from \cite{XENON:2024ijk} for XENONnT. 
We use a total of three bins for XENONnT of size (90,160], (160,230], (230,500] PE and eight bins of equal width between [4,8) electrons for PandaX-4T.
We find an excellent agreement for the SM event rates for PandaX-4T and a slightly worse but still acceptable agreement for XENONnT
which may be due to details in the experimental analysis.

We implement the following $\chi^2$ function with nuisance parameters for signal and background $\alpha,~\beta$ in our analysis
\begin{align}
\chi^2(\alpha, \beta) &= 2 \sum_i \left[ \hat{N}_i - N_i^{\text{meas}} + N_i^{\text{meas}} \log \left( \frac{N_i^{\text{meas}}}{\hat{N}_i} \right) \right]\nonumber \\
&\quad + \left( \frac{\alpha}{\sigma_\alpha} \right)^2 + \left( \frac{\beta}{\sigma_\beta} \right)^2 \\
\hat{N}_i &= N_i^{\text{model}} (1 + \alpha) + N_i^{\text{bkg}} (1 + \beta) 
\label{chi2binned}
\end{align}
where $N_i^{\text{model}}$, $N_i^{\text{bkg}}$, $N_i^{\text{meas}}$ are the number of predicted signal events in the model, background events, and experimentally measured events in the $i^{th}$ bin respectively. The systematic uncertainties of the signal and background $\sigma_\alpha$ and $\sigma_\beta$ are 
  $\sigma_\alpha = 0.17 ~(0.22)$ and $\sigma_\beta = 0.20 ~(0.05)$ for PandaX-4T (XENONnT) \cite{PandaX:2024muv, XENON:2024ijk}.

With this approach our results for 
 equal coupling of a new vector mediator to all three flavors and equal coupling to up and down quarks  agrees with the results from \cite{Blanco-Mas:2024ale,DeRomeri:2024hvc}.

\section{Results}
\label{sec:results}
\subsection{LMA-Dark in muon and tau neutrino sector}
\begin{figure}
    \centering
    \includegraphics[width=\linewidth]{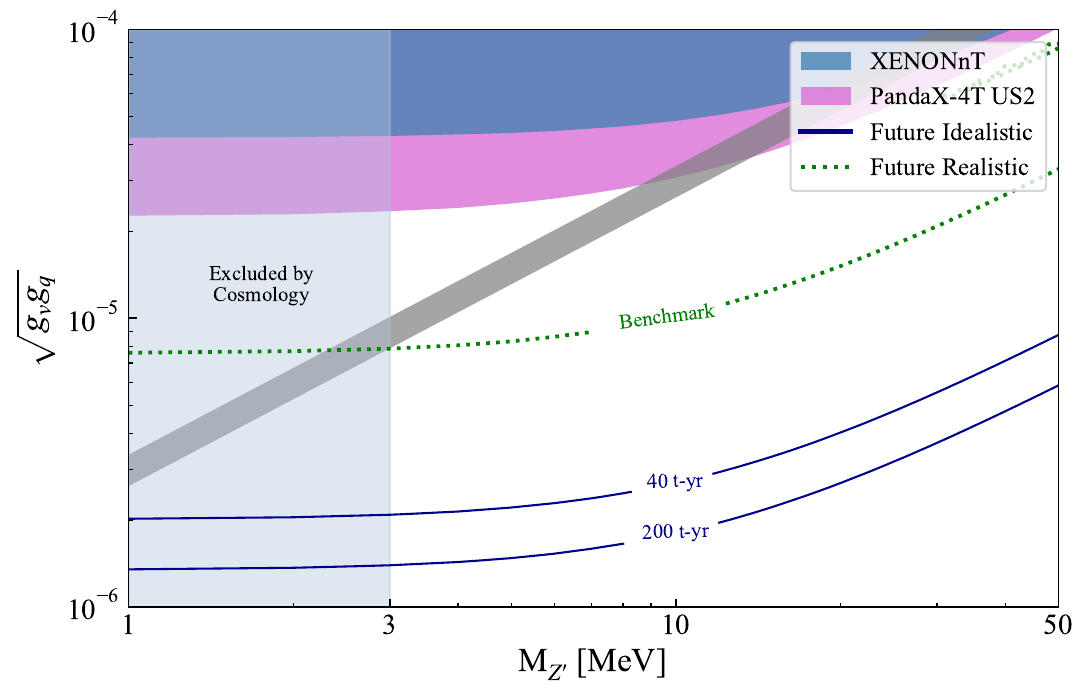}
    \caption{Constraints on the coupling strength of a new light mediator with mass $m_{Z'}$ to quarks and muon and tau neutrinos $\sqrt{g_\nu g_q}$ at
    3$\sigma$ C.L.   for 2 d.o.f. from the XENONnT and PandaX-4T data \cite{XENON:2024ijk, PandaX:2024muv}.  We assume equal  couplings to muon and tau neutrinos and equal couplings to up and down quarks. The grey band corresponds to the  LMA-Dark region from oscillation data from \cite{Esteban:2018ppq} drawn at $3\sigma$ for 1 d.o.f. The region to the left is excluded from cosmology at $2\sigma$ (1 d.o.f.) \cite{Sabti:2021reh}. The green dashed line corresponds to the required exposure of 125 t-yr exposure Xe experiments assuming all other experimental characteristics  are similar to PandaX-4T. The solid blue lines show the expected sensitivity of an ideal xenon experiment without background and 100\% signal efficiency with an exposure of 40 t-yr and 200 t-yr. }
    \label{fig:lmadmmtt}
\end{figure}

We turn now to our results. 
We show in fig.~\ref{fig:lmadmmtt}
the constraints derived from the XENONnT and PandaX-4T data sets \cite{XENON:2024ijk, PandaX:2024muv} in the $m_{Z'}-\sqrt{g_\nu g_q}$  plane for a mediator with equal couplings to muon and tau neutrinos and equal couplings to up and down quarks compared to the LMA-Dark region from \cite{Esteban:2018ppq}.
The data from current experiments only excludes the LMA-Dark region above  mediator masses of $m_{Z'}\gtrsim 10 $ MeV.

As current data does not fully exclude LMA-Dark in the muon and tau sector, we  consider possible future DM direct detection experiments and test their sensitivity. 
PandaX-4T \cite{PANDA-X:2024dlo} is proposed to be  upgraded to the PandaX-xT experiment, and will feature a 43 t liquid xenon target. The experiment is expected to conservatively have an efficiency as good as the current generation PandaX-4T \cite{PANDA-X:2024dlo}, with the possibility of improved efficiency and lower backgrounds. XLZD is the planned upgraded detector of the XENON and LZ  experiments \cite{DARWIN:2016hyl, XLZD:2024gxx}. In the conservative plan, XLZD aims to have a 60 t liquid xenon target. Both future xenon experiments expect to reach a 200 t-yr exposure by the end of their lifetimes. Darkside-20k \cite{DarkSide-20k:2017zyg} and ARGO  \cite{argo} are  planned DM experiments which would feature a 48 t and a 300 t liquid argon target and they are expected to achieve a threshold of 0.5 keV$_{nr}$, when utilizing the ionization only channel \cite{Darkside_Cevns}.
The number of CEvNS events in future xenon and argon detectors is similar, due to the similar nuclear masses, and we therefore focus on the xenon experiments in the following but similar results also apply to future argon experiments.

To simulate future experiments, we use the following two scenarios: In the idealistic scenario we increase the exposure of PandaX-4T to 40 t-yr and 200 t-yr, corresponding to expected exposures of Panda-xT and XLZD and   we assume 100$\%$ efficiency, treat the experiment as a counting experiment, and we assume  no background. We assume a threshold of 0.1 keV$_{\text{nr}}$, lower than current generation experiments. 
Alternatively, we consider a realistic experimental setup where we keep the systematic uncertainty  and signal to background ratio fixed to the values of PandaX-4T. We use $\sigma_\alpha=0.2$, $\sigma_\beta=0.2$ and the signal to background ratio  varies across the bins between 0.25 to 0.46. 
We maintain the same number of bins as PandaX-4T in modeling the realistic future scenario. Our sensitivity benefits from the energy spectrum of the data across the 8 bins and we find that it drastically falls if halving the number of bins.

From fig.~\ref{fig:lmadmmtt}, in the idealistic scenario with 40 t-yr or 200 t-yr the LMA-Dark region can be easily excluded at more than 5$\sigma$.  
For a realistic setup we find that 125 t-yr of exposure with the current experimental parameters is required to rule out LMA-Dark at 3$\sigma$. The current exposures are 3.51 t-yr for XENONnT and 1.04 t-yr for PandaX-4T.

\begin{figure}
    \centering
    \includegraphics[width=\linewidth]{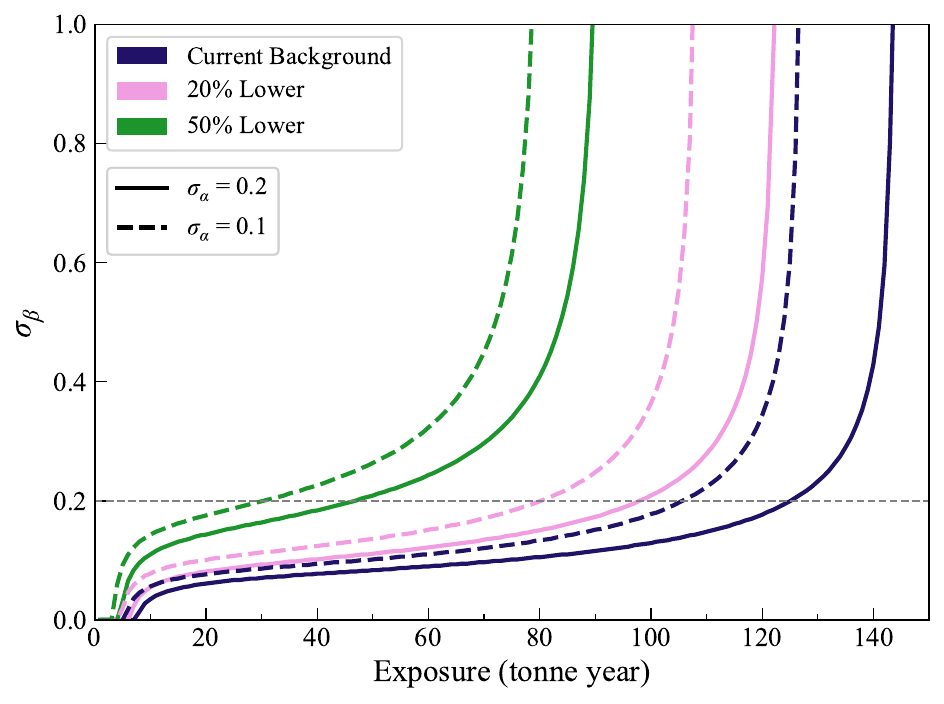}
    \caption{Required systematic uncertainties on signal $\sigma_\alpha$ and background  $\sigma_\beta$   to rule out the LMA-Dark point with $m_{Z'}=3$ MeV and $\sqrt{g_\nu g_q}=10^{-5}$ at 3$\sigma$ (1 d.o.f.) as a function of exposure. Current experiments have $\sigma_\alpha\approx0.2$ and $\sigma_\beta\approx 0.2$.  We assume either the current background  or a reduced background of 20\% or 50\%. The current exposures are 3.51 t-yr for XENONnT and 1.04 t-yr for PandaX-4T. See fig.~\ref{fig:5sigbenchmark} for the results to rule out LMA-Dark at $5\sigma$. }
    \label{fig:3sigbenchmark}
\end{figure}

If the experimental parameters improve in future iterations of DM direct detection experiments LMA-Dark can be ruled out with lower exposure. In fig.~\ref{fig:3sigbenchmark} we show the required values of $\sigma_\alpha$ and $\sigma_\beta$ as  functions of exposure which excludes LMA-D at $3\sigma$ (see appendix \ref{sec:5sigma} for the results at $5\sigma$). 
Depending on the background level and the systematic uncertainty on the signal and background events, the required exposure can be reduced to $\mathcal{O}(10~\text{t-yr})$.
Lowering the background requires reducing the radioactivity on
the cathode electrode which dominates the number of background events for the
 US2 data of PandaX-4T \cite{PandaX:2024muv} or to reduce the  accidental  coincidence  events for XENONnT \cite{XENON:2024ijk}.

Given that the current signal-to-background ratio of the total number of events is 0.3 for PandaX-4T with varying signal-to-background ratios across the bins between 0.25 to 0.46, decreasing the systematic uncertainty on the signal $\sigma_\alpha$ is less impactful than decreasing the uncertainty on the background $\sigma_\beta$.
The systematic uncertainty of the signal comes from the experimental data selection and the $^8$B solar neutrino flux. The latter uncertainty could potentially be reduced in the future as upcoming experiments, in particular solar neutrino measurements at DUNE \cite{Capozzi:2018dat,DUNE:2020ypp,Meighen-Berger:2024xbx},   can measure the $^8$B flux with an improved precision of 1.6\% \cite{Meighen-Berger:2024xbx} compared to the current uncertainty of 5\% \cite{SNO:2011hxd,Super-Kamiokande:2023jbt,BOREXINO:2018ohr}.

\begin{figure}
    \centering
    \includegraphics[width=\linewidth]{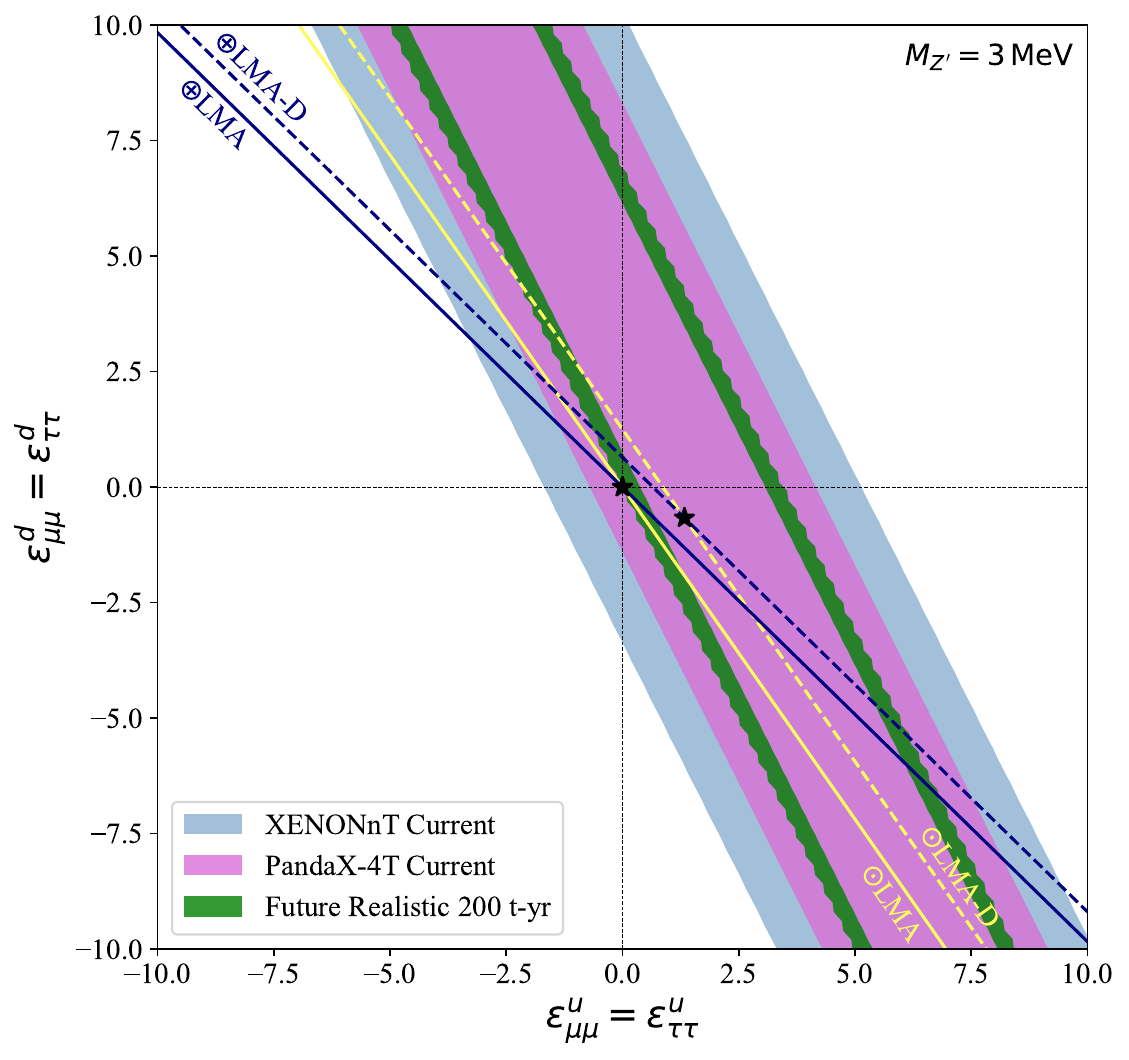}
    \caption{Allowed regions at 3$\sigma$  (1 d.o.f.) from the XENONnT and PandaX-4T data \cite{XENON:2024ijk, PandaX:2024muv} for a light mediator with $m_{Z'}=3$ MeV with equal couplings to muon and tau neutrinos as a function of its coupling to up-and down quarks at a momentum transfer of 10 MeV. We also show the expected constraints from a future xenon experiment with 200 t-yr exposure. We also show the LMA and LMA-Dark region using the matter densities of the Earth and
the Sun. These lines intersect at the stars which means that no oscillation probe can distinguish these cases.  }
    \label{fig:lmadheavy}
\end{figure}

If the assumption of equal couplings to up and down quarks is softened, the excluded regions for a mediator of mass $m_{Z'}=3$ MeV coupling with equal couplings to  muon and tau neutrinos is shown fig.~\ref{fig:lmadheavy}. A future xenon experiment with 200 t-yr exposure will narrow the allowed regions drastically.
The slope of the bands in the $\varepsilon_{\mu\mu,\tau\tau}^u-\varepsilon_{\mu\mu,\tau\tau}^d$ plane is given by $(2N+Z)/(2Z+N)$. 
Therefore, materials with a different neutron to proton ratios are required to break the degeneracies in the $\varepsilon^u_{\alpha\alpha}-\varepsilon^d_{\alpha\alpha}$ parameter space and shrink the allowed regions. 
Note that oscillation data using neutrino propagation through the Earth and the Sun only slightly narrow down this parameter space \cite{Denton:2022nol} as the neutron to proton ratios  $Y_n=N/Z$ differ only by a factor of $\sim 3$ between 1.05 for the Earth and 0.3 in the Sun for the peak of the  $^8$B flux
\cite{Bahcall:2004pz} compared to $Y_n \approx 1.4$ of Xenon.  We show in fig.~\ref{fig:lmadheavy} the LMA and LMA-Dark degenerate regions in the Sun and in the Earth. Note that there is a degenerate point which cannot be  distinguished from  the case of no new physics  in oscillation experiments.

\subsection{Distinguishing new physics scenarios in the muon and tau neutrino sectors}
DM experiments observing CEvNS with solar neutrinos are uniquely positioned to test  new interactions in the muon and tau sector  and distinguish them from new interactions in the muon or tau sector only. 
Apart from the LMA-Dark degeneracy such new interactions could also arise for example in $L_\mu-L_
\tau$ models \cite{Ma:2001md,He:1991qd,He:1990pn}, see \cite{Amaral:2021rzw} for a study on probing such models with neutrinos.

\begin{figure}[t]
    \centering
    \includegraphics[width=\linewidth]{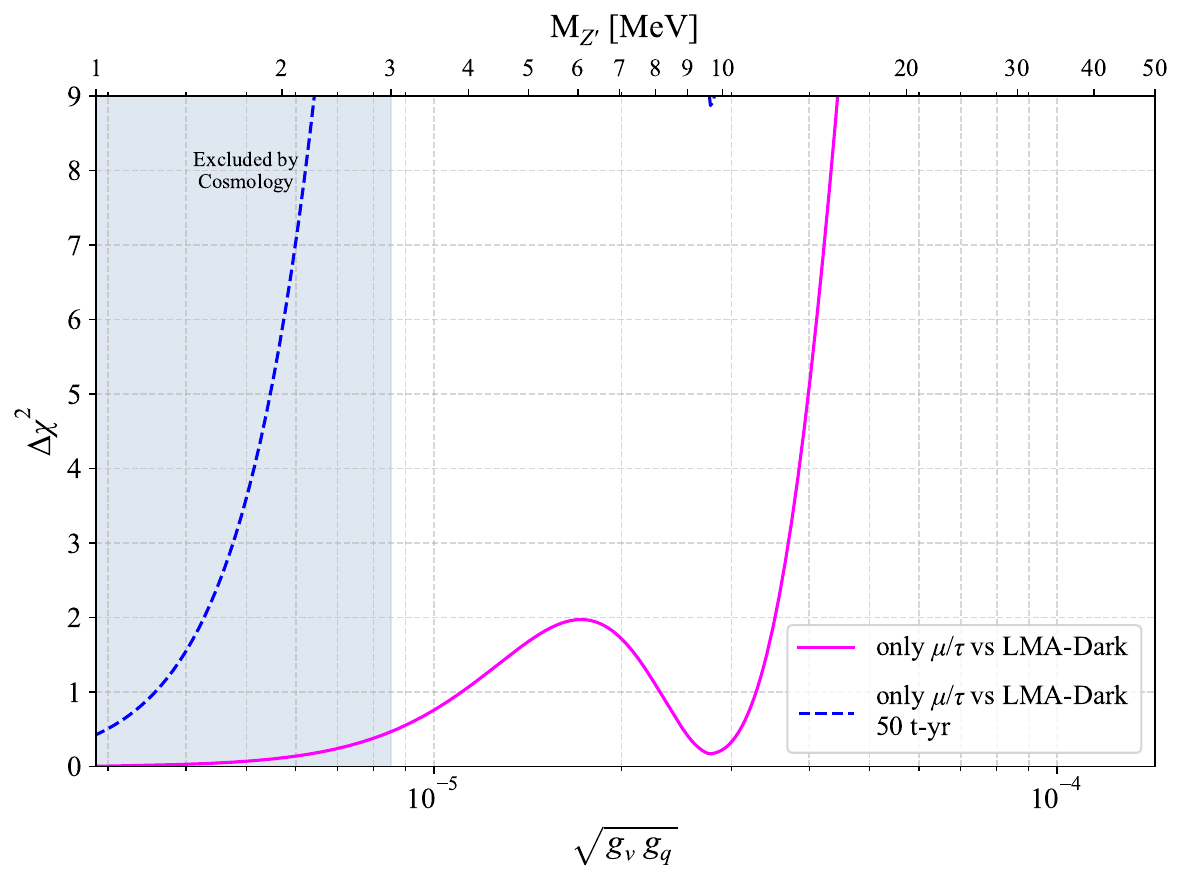}
    \caption{Distinction power of a PandaX-4T-like  experiment between the LMA-Dark scenario and a model where a new mediator couples with a LMA-Dark-like-strength to only muon or only tau neutrinos. Since we assumed $\sin^2\theta_{23}=0.50$ and $\cos\delta=0$ the transition probabilities $P_{e\mu}$ and $P_{e\tau}$ are identical and the results for a mediator coupling either only to muon neutrinos or only tau neutrinos are the same. The solid lines show the result using current PandaX-4T data, while the dashed lines assume an increased exposure of 50 t-yr. } 
    \label{fig:dist}
\end{figure}

To illustrate the power of DM experiments to disentangle a new light mediator coupling to muon and tau neutrinos or just one flavor, we assume that LMA-Dark is realized in nature with a light mediator and a coupling corresponding to the LMA-Dark region with  $
g_{\nu_\alpha} g_f = \frac{\sqrt{2}}{2} \, G_F M_{Z'}^2 $. We test then a scenario where the light mediator couples to only  muon or  only tau neutrinos with the coupling strength corresponding to the LMA-Dark region. Since we assumed $\sin^2 \theta_{23}=0.50$ and $\cos\delta=0$ the transition probabilities $P_{e\mu}$ and $P_{e\tau}$ are identical leading to the same results for a mediator  coupling to either only muon neutrinos or only tau neutrinos. The distinction  power between these two scenarios  parametrized as  $\Delta\chi^2=\chi^2(\mu~\text{ or } \tau)-\chi^2(\text{LMAD})$ is shown in 
fig.~\ref{fig:dist}. We see that current data can only distinguish between these scenarios for mediator masses above 15 MeV at $2\sigma$. 
An exposure of 50 t-yr for a PandaX-4T-like experiment is required to distinguish  between these scenarios at $>3\sigma$ down to mediator masses excluded by cosmology.
Around a mediator mass of  $m_{Z'}\approx 10 $ MeV, 
the number of events in our scenarios is close to the SM value, making it challenging to exclude new physics and leading to a reduced sensitivity.

\subsection{LMA-Dark in the electron neutrino sector}

\begin{figure}
    \centering
    \includegraphics[width=\linewidth]{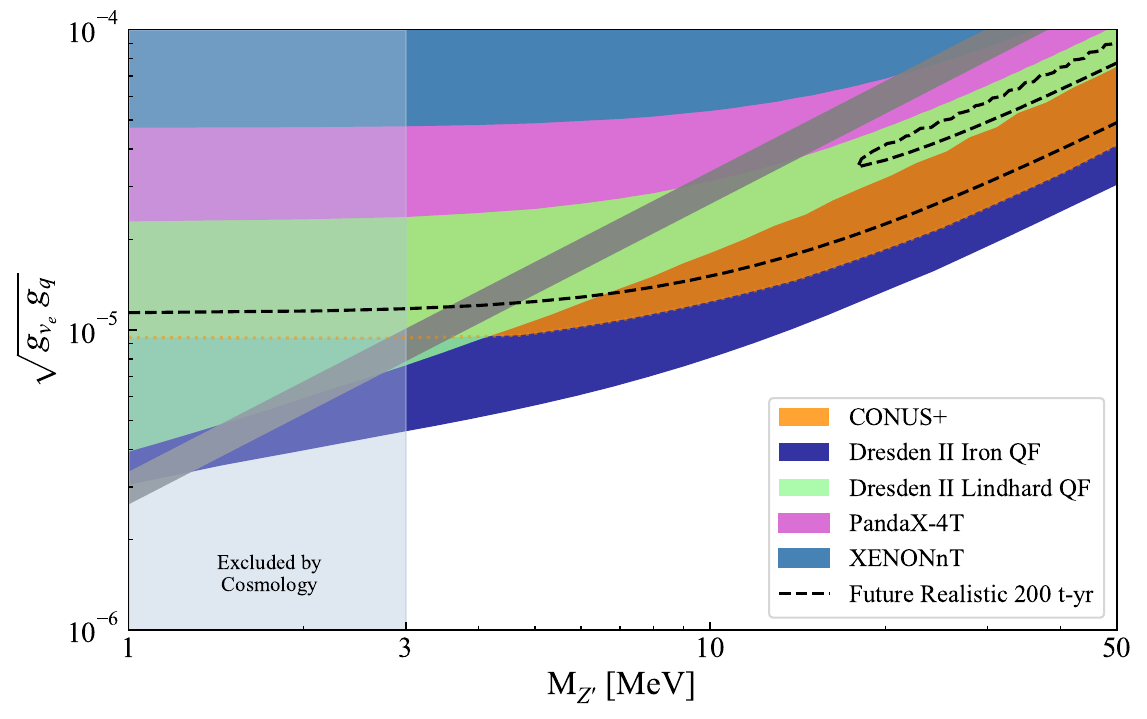}
    \caption{The constraints at 3$\sigma$ C.L.  (2 d.o.f.)  for light mediator coupling to electron neutrinos only with equal couplings to up and down quarks from the XENONnT and PANDAX-4T data \cite{XENON:2024ijk, PandaX:2024muv}.
   We also show the constraints from the observation of CEvNS with reactor neutrinos from CONUS+ at 90\% C.L. from \cite{Chattaraj:2025fvx,DeRomeri:2025csu} and from Dresden-II at $3\sigma$ for two different choices of quenching factors from \cite{AristizabalSierra:2022axl, Denton:2022nol}.
  We also show the forecasted sensitivity of 200 t-yr of a PandaX-4T-like experiment. }
    \label{fig:lmadee}
\end{figure}

If LMA-Dark is realized in the electron neutrino sector  with couplings to quarks, i.~e.~$\eps_{ee}=-2$ and all other NSI parameters are zero, DM experiments can probe this region as well. 
 Recently, the first observations of CEvNS with reactor neutrinos has been announced from the Dresden-II experiment \cite{Colaresi:2022obx,Colaresi:2021kus} and CONUS+
\cite{Ackermann:2025obx} which were analyzed in
\cite{Coloma:2022avw,AristizabalSierra:2022axl,Liao:2022hno,Chattaraj:2025fvx,DeRomeri:2025csu,AtzoriCorona:2025ygn}. We show their constraints in fig.~\ref{fig:lmadee} together with the current constraints from PandaX-4T and XENONnT.  While current XENONnT and PandaX-4T data only probes the LMA-Dark region down to mediator masses of 10 MeV, similar to LMA-Dark in the muon and tau sector due to the similar neutrino flux, data from the Dresden-II experiment fully excluded the region, independent of the choice of quenching factor and the CONUS+ results cover the vast majority of parameter space as well. 
Future PandaX-4T-like experiments  can provide complementary probes to reactor CEvNS experiments but require an exposure of at least 600  t-yr to exclude LMA-Dark at $3\sigma$.

\section{Conclusions}
\label{sec:conclusion}
The first observation of the neutrino background at DM direct detection experiments 
allows to probe new contributions to the  neutrino scattering cross section at low energies. 
 Of particular interest is a region of parameter space of new neutrino interactions which leads to the LMA-Dark degeneracy in neutrino oscillations.  This degeneracy of the oscillation Hamiltonian corresponds to a change in the mass ordering in the presence of new vectorial neutrino interactions with strength of order  the Fermi constant. Here we have shown that DM direct detection experiments have an unparalleled possibility to probe a yet unexplored region of LMA-Dark parameter space, namely equal couplings of a new light mediator to muon and tau neutrinos. This is due to the presence of all three neutrino flavors in the $^8$B solar neutrino flux coming from the transition probabilities of electron neutrinos to muon and tau neutrinos in addition to the non-vanishing electron neutrino survival probability \cite{Wolfenstein:1977ue} and their subsequent detection in the low-energy CEvNS scattering process.
 
 We have shown that current solar neutrino CEvNS data from XENONnT and PandaX-4T does not yet fully exclude the LMA-Dark parameter space for equal couplings to muon and tau neutrinos, however, assuming similar experimental details, xenon-based DM experiments with an exposure of 125 t-yr are sufficient to achieve this goal at 3$\sigma$. 
  Finally, we have also shown that DM experiments provide a complementary way to probe LMA-Dark in the electron sector, that is a light mediator coupling to quarks and electron neutrinos, compared to reactor neutrino CEvNS experiments. 

The remaining LMA-Dark regions of parameter space are equal interactions of muon and tau neutrinos with electrons mediated by a low mass vector boson and can  be probed in the future in large-scale solar neutrino-electron scattering experiments like JUNO, DUNE, or HyperKamiokande or DM direct detection experiments.

\section*{Acknowledgements}
We thank Peter Denton for  useful communication concerning this work.
This project has received support by
the U.~S.~ Department of Energy Office of Science under award number DE-SC0025448.

\appendix

\begin{figure}
    \centering
    \includegraphics[width=\linewidth]{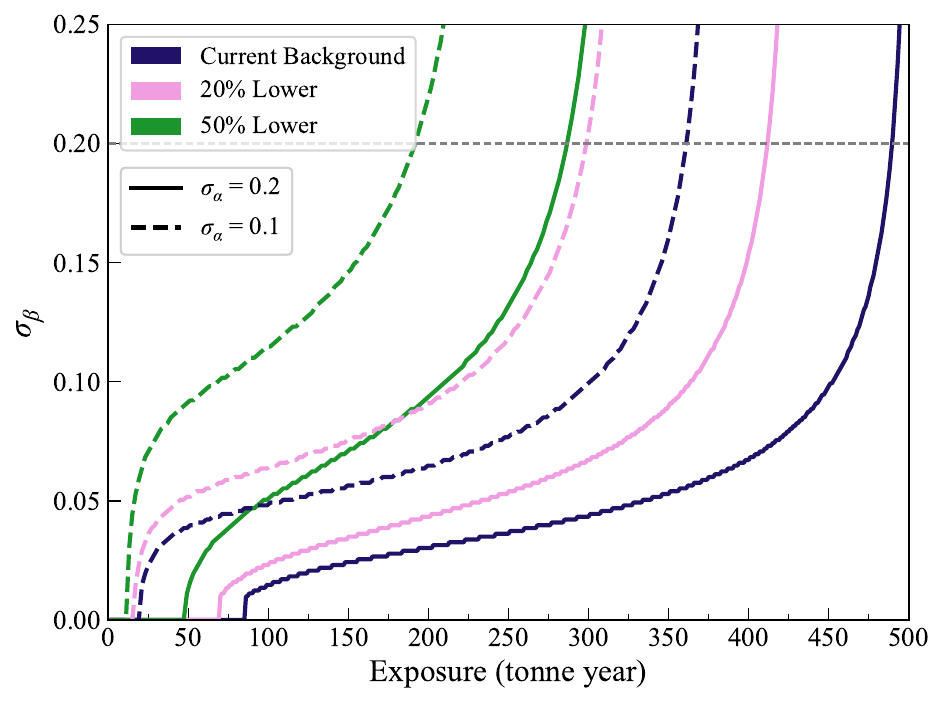}
    \caption{ 
    Required systematic uncertainties on the background  $\sigma_\beta$  and signal $\sigma_\alpha$ required to rule out the LMA-D point with $m_{Z'}=3$ MeV and $\sqrt{g_\nu g_q}=10^{-5}$ at 5$\sigma$ (1 d.o.f.) as a function of exposure. Current experiments have $\sigma_\alpha=0.2$ and $\sigma_\beta=0.2$.  We assume either the current background  or a reduced background of 20\% or 50\%. The current exposures are 3.51 t-yr for XENONnT and 1.04 t-yr for PandaX-4T. See fig.~\ref{fig:3sigbenchmark} for the results to rule out LMA-Dark at $3\sigma$.}
    \label{fig:5sigbenchmark}
\end{figure}
\section{Results at $5\sigma$
}
\label{sec:5sigma}
In fig.~\ref{fig:5sigbenchmark} we show the required systematic uncertainties to exclude LMA-Dark in the muon and tau sector at $5\sigma$, compare to fig.~\ref{fig:3sigbenchmark} in the main text which shows the same results at $3\sigma$.

\bibliography{main}

\end{document}